# Optimization of Mechanical Design Bladeless Wind Turbine for Electricity Fulfilment in Nusa Tenggara Timur, Indonesia

Muhammad Farhan Ramadhany ( ✉ farhanramadhany@mail.ugm.ac.id )
  Department of Nuclear Engineering and Engineering Physics, Faculty of Engineering, Gadjah Mada University   https://orcid.org/0000-0003-3183-9063

Theo Aden Kusuma
  Department of Nuclear Engineering and Engineering Physics, Faculty of Engineering, Gadjah Mada University

Yessika Natalia Chelsie
  Department of Nuclear Engineering and Engineering Physics, Faculty of Engineering, Gadjah Mada University

Gandha Satria Adi
  Department of Nuclear Engineering and Engineering Physics, Faculty of Engineering, Gadjah Mada University

# Optimization of Mechanical Design Bladeless Wind Turbine for Electricity Fulfilment in Nusa Tenggara Timur, Indonesia

Muhammad Farhan Ramadhany[a*], Theo Aden Kusuma[a], Yessika Natalia Chelsie[a], Gandha Satria Adi[a]

[a] *Department of Nuclear Engineering and Engineering Physics, Faculty of Engineering of Universitas Gadjah Mada, Republic of Indonesia*

**Abstract** The current government's goal is the realization of an even distribution of the electrification ratio in Indonesia. However, in 2018 the electrification ratio in Nusa Tenggara Timur (NTT) only reached 62.07%. One of the solutions offered is the implementation of a Bladeless Wind Turbine (BWT). BWT is a type of wind power plant that can work optimally in areas with low wind speeds, 3 to 8 m/s, which is still above the NTT average wind speed, 2.3 m/s. This study aims to optimize the mechanical design of BWT with shape and size parameters based on the manipulation of the coefficient of friction through computational fluid dynamic simulations that can work optimally according to wind speed in NTT. In this study, 3 design variations were used, namely Initial Design, Modified Design 1, and Modified Design 2. Based on the research that has been done, MD2 has better results than MD1 and ID, and it can be said that MD2 can work optimally in wind speed areas. low. So MD2 is a mechanical design model that is very suitable to be applied in NTT and is expected to be a recommendation in making masts for BWT-based wind power plants in Indonesia.

## 1. Introduction

Electricity is a primary need for world civilization. The need for electricity continues to increase along with the times, this is evidenced by almost all equipment and human activities using electrical energy as the main energy source. Thus, it can be said that electricity has become a very important thing for the world community.

Currently, the Indonesian government is trying to achieve equal distribution of electrification throughout Indonesia. However, until 2018 the electrification ratio in NTT only reached 62.07% or more than 3 million people in NTT still cannot use electricity (Kementrian ESDM, 2018). One solution that has been done is the application of wind power plant. However, conventional wind power plant currently in use can only achieve optimal conditions in areas with average wind speeds of 10 to d. 12 m/s, far from the average wind speed in the NTT region which is only 2.3 m/s. Based on this, an innovative solution is needed, namely the application of wind power plant based on Bladeless Wind Turbine (BWT) which is optimized by mechanical design. BWT is a power plant with high efficiency in areas with low wind speed. The electrical energy converted from kinetic energy is caused by the Von Karman Vortex Street phenomenon which causes resonance and oscillation in the BWT mast. However, BWT can operate optimally in the low wind speed range, which is around 3 to 8 m/s. Therefore, it is necessary to optimize the mechanical design of the BWT mast so that it can operate optimally in NTT where the average wind speed is only 2.3 m/s to achieve an even distribution of the electrification ratio in NTT.

## 2. Literature Review

### 2.1. Bladeless Wind Turbine

BWT is one of the high efficient wind power plants in areas with low wind speeds (Asre et al., 2021). In contrast to conventional wind power plant which uses propellers to generate electrical energy based on a certain wind speed and direction, BWT utilizes low wind speeds that move from all directions to generate electrical energy without using propellers (Balakrishnan et al., 2019).

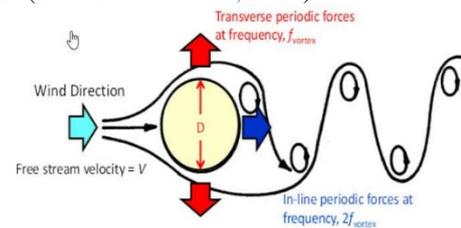

**Figure 1.** The phenomenon of vortices through the cylinder will produce an oscillating wake characteristic

In the working principle of BWT, energy conversion occurs in the mast section, where the wind hits the mast and makes it vibrate at the same frequency (Fan, Wang, & Tan, 2021). This induced resonance phenomenon is referred to as the Von Karman Vortex Street Effect or vortex shedding (Fig. 1). Vortex shedding is an oscillating flow due to a fluid such as air or water flowing and passing through the mast surface (Thomai et al., 2019).

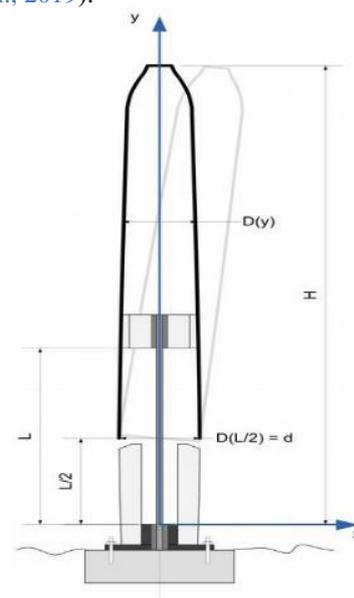

**Figure 2.** Bladeless wind turbine section view (Yáñez Villarreal, 2018)

---

* Corresponding author
  e-mail: farhanramadhany@mail.ugm.ac.id (M. F. Ramadhany)

When the fluid hits the projected mast surface area from a wind direction, the streamlines of the fluid will tend to scatter and form a turbulent flow. This is what results in the formation of fluid currents called vortices (Francis et al., 2021). Mast exposed to the wind will tend to oscillate because of the vortices formed in the mast structure and cause linear oscillatory motion which is then converted into rotational motion to produce electrical energy (Fig. 2) (Abhijit Mane1, 2017; Tandel et al., 2021).

In terms of cost-effectiveness, one of the advantages of BWT is its low cost. Based on the Levelized Cost of Energy (LCOE), BWT will have a low category LCOE and a faster return on investment. Cost reductions can come from the design, use of raw materials, and the unnecessary need for conventional nacelles, support mechanisms, propellers, and wind generators (Manshadi et al., 2021; Tandel et al., 2021). In addition, the simple design of the BWT without the propeller means that lubricating oil is not used so that it can reduce waste. In addition, unlike conventional wind power plants which can disturb wildlife due to the large propellers and noise pollution produced, BWTs without propellers do not harm the surrounding wildlife and the sound produced is so small that it is not disturbing (Yáñez Villarreal, 2018).

*2.2. Electrification Ratio*

The electrification ratio is the percentage ratio of the number of electrified households divided by the number of national households. The number of electrified households comes from the number of PLN households and the number of non-PLN households (Kementrian ESDM, 2018). According to the Electricity Statistical Data by the Ministry of Energy and Mineral Resources, in 2018 the national electrification ratio increased by 2.95% from 2017. This data shows that the reach of electricity facilities cannot be "enjoyed" by all people in Indonesia. The former minister of ESDM, Ignasius Jonan targets the electrification ratio in Indonesia to reaching 99% by the end of 2019 (Marroli, 2019). The fulfilment of the electrification ratio to reach 100%, which is demonstrated by the reach of electricity facilities to remote parts of the country. Several areas, such as NTT, have become one of the main focuses for the development of electricity facilities. This is because, according to the Electricity Statistical Data by the Ministry of Energy and Mineral Resources, the electrification ratio of NTT Province is only 62.07% or more than 3 million people in NTT who still do not get electricity facilities.

*2.3. Wind Condition*

The geographical location of Indonesia, which passes through the equator, makes Indonesia generally have relatively low wind speeds. According to data from BPS, in 2015, the average wind speed in Indonesia only reached 2.42 m/s with a maximum wind speed of only 4.05 m/s and a minimum of 0.07 m/s. In NTT Province, the average wind speed from the Kupang Climatology Station and Eltari Meteorological Station in February 2020 was 2.3 m/s. In addition, the geographical location of NTT also consists of several mountains and coasts which cause a lot of local winds to blow. The local wind is caused by the difference in heat. The difference in heat between day and night results in a heat difference between the air and land (land winds) and air and ocean (sea breezes). In addition, the heat difference also occurs between air with highlands (mountain winds) and air with lowlands (valley winds) (Tjasyono, 2008).

*2.4. Coefficient of Side Force*

Lift coefficient (CL) and drag coefficient (CD) are dimensionless parameters that determine the lift and drag force of a tool, where drag is the force generated from the direction of wind speed and lift is normal force generated at wind speed (Fan, Wang, Wang, et al., 2021; Munson et al., 2013). The drag coefficient is closely related to the geometry or mechanical design of a tool because the size of the drag coefficient value indicates the amount of resistance produced by the fluid, namely the wind speed on the mast. The ratio of lift coefficient (CL) to drag coefficient (CD) is the coefficient of oscillatory drift or gliding ratio. Where the greater the ratio CL/CD indicates greater the coefficient of oscillatory deviation experienced by the model. The magnitude of the coefficient of the oscillatory displacement force can be used as the basis for a more efficient mechanical design review to be applied because the larger the swipe area or deviation, the more maximal power and vortex shedding it will be. Therefore, optimization of the mechanical design of the mast can be done by reducing the drag coefficient and increasing the lift coefficient of the mast (Federal Aviation Administration, 2013).

## 3. Materials and Methods

In completing this research, the types of research carried out are literature studies and Computational Fluid Dynamics. Literature study, namely the method used as a theoretical basis in solving scientific problems. This method is carried out by using several journals, books, and statistical data as a reference in the form of writing. In addition, the Computational Fluid Dynamics method is a computer simulation that begins with making an operational model of the system according to the existing problems, then simulating it to obtain a fluid flow model that resembles the actual conditions (Maulana, 2016). The stages of the research carried out are as follows:

*3.1. Dimensional Analysis*

Optimizing the BWT design so that it can work well at wind speeds lower than the ideal working range of the BWT, dimensional analysis is needed to determine the variables that can affect the oscillation ability of the BWT. One of the influential variables is the lift coefficient (CL) and drag coefficient (CD), to optimize the performance of the BWT mechanical design changes to manipulate the values of CL and CD that work on the BWT.

*3.2. Bladeless Wind Turbine Modelling*

Based on the dimensional analysis, to model the optimal BWT design, a large coefficient of friction force (CL/CD) is required. In other words, BWT modelling is carried out by reducing the contact area of the BWT mast which is in the direction of the incoming wind speed. Therefore, this study used three forms of design variations as follows:
1. Initial Design, which is a BWT design in general in the form of a cylinder with a diameter of 10 cm.
2. Modified Design 1, which is a variation of the BWT design in the form of a 5 cm diameter cylinder.

3. Modified Design 2, which is a variation of the BWT design in the form of an ellipse with a short diameter of 5 cm and a long diameter of 15 cm.

*3.3. Simulation Model Formulation*

The simulation model formulation stage uses Autodesk Inventor and simFlow 3.1 software. The simulation model made must be following with the mathematical equations, parameters, and determination of initial conditions. In the formulation of the design model by making a mast structure which will then be simulated. The simulation parameters used are as follows:
1. Constant kinematic viscosity of 1.50E-05 m2/s
2. Air is a Newtonian fluid
3. Flown inlet from $x^-$ axis and outlet $x^+$ axis
4. Boundary Conditions
    a. Inlet as a certain incoming air velocity (u)
    b. outlet as out pressure
5. Initial Conditions
    a. zero initial pressure
    b. The wind speed for a certain height follows the empirical equation (1)
    $$\frac{0,5 \times U}{0,1} \times (y + 0,05) \qquad (1)$$
    where U is the wind speed (m/s), and y is the height (m).

After the above parameters are defined, a simulation is carried out by iterating the BWT design model with variations in wind speed treatment from 1 to 5 m/s.

*3.4. Data Analysis*

The data analysis is carried out by making comparisons based on animation, frequency graphs, and gliding ratio graphs from CFD simulation results to determine the design that works best based on a predetermined speed.

## 4. Result and Discussion

In this study, the design of the BWT mast that can work optimally in areas with low wind speeds (1 m/s to 5 m/s) was determined by doing a comparison between the Initial Design (ID) simulation results with Modified Design 1 (MD1) and Modified Design. Design 2 (MD2), to get a model that has an optimal tendency to oscillate with a predetermined wind speed. The oscillation tendency will be discussed based on the comparison of several parameters as follows: fluid motion analysis parameters based on animation are used to determine the minimum speed for the BWT model to oscillate; frequency parameters to get the value of the oscillation frequency of each BWT model; as well as the comparison of the lift constant ratio (CL) with the drag constant (CD) parameters to show the magnitude of the coefficient of oscillatory displacement at BWT.

Based on the results of the CFD simulation, which is presented in Fig. 3, 4, and 5 performed on each model shape with variations in speed, the shape of the airflow stability after passing through the mast can be obtained. The stability of the air after passing through the object that has been obtained can show the stability of BWT oscillations when applied. It can be seen in the Initial Design (ID) animation showing the stability of the air flow after passing through the object when the wind speed is 1 m/s to 5 m/s as shown in Table 1. Furthermore, Modified Design 1 (MD1) shows the stability of the air flow after passing through the object when the wind speed is 1 m/s to 5 m/s. While the Modified Design 2 (MD2) shows the stability of the air flow after passing through the object when the wind speed is 1 m/s to 5 m/s. This shows that the ID model will start to oscillate at a wind speed of 3 m/s, while the MD1 and MD2 models will start to oscillate at a wind speed of 1 m/s. So, it can be seen that the BWT with the MD1 and MD2 models can already oscillate in an area with a wind speed of 1 m/s.

In addition, this is supported by the Coefficient Force graph from the simulation results that have been carried out with 3 Bladeless Wind Turbine (BWT) mechanical models, which show the frequency, lift coefficient (CL), and drag coefficient (CD) in each model in the Table 1.

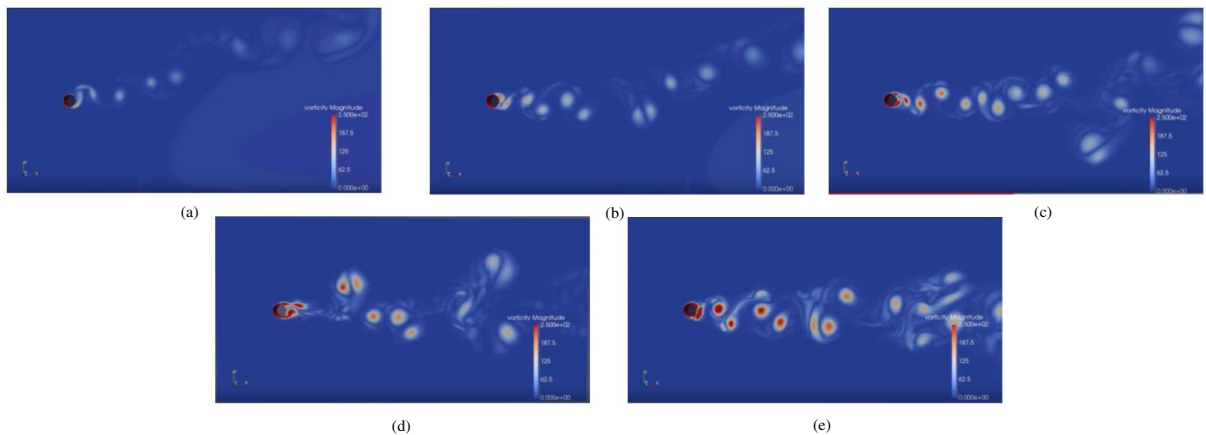

**Figure 3.** Initial design vortices at 1 m/s (a), 2 m/s (b), 3 m/s (c), 4 m/s (d), and 5 m/s (e) of wind speed

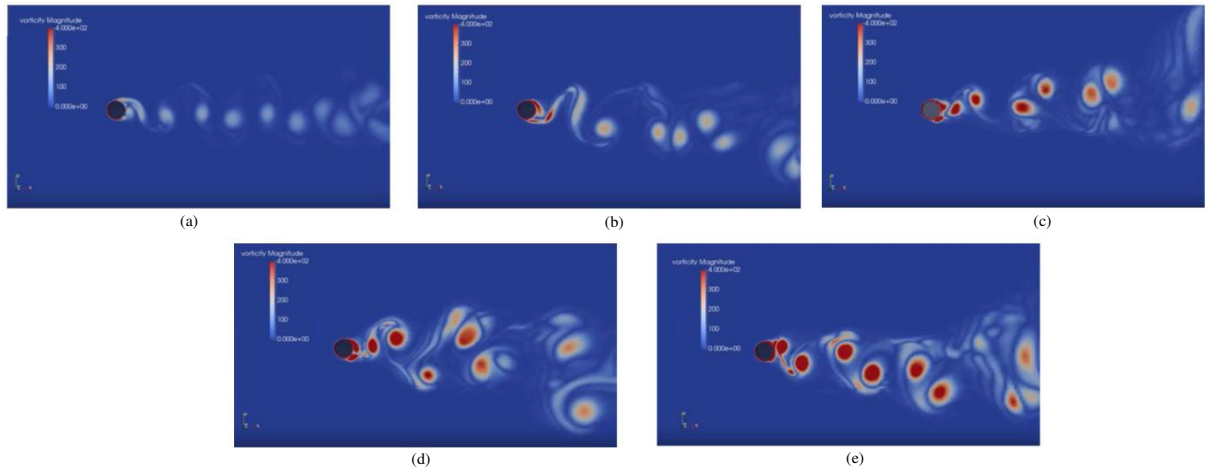

**Figure 4.** Vorticity modified design 1 at 1 m/s (a), 2 m/s (b), 3 m/s (c), 4 m/s (d), and 5 m/s (e) of wind speed

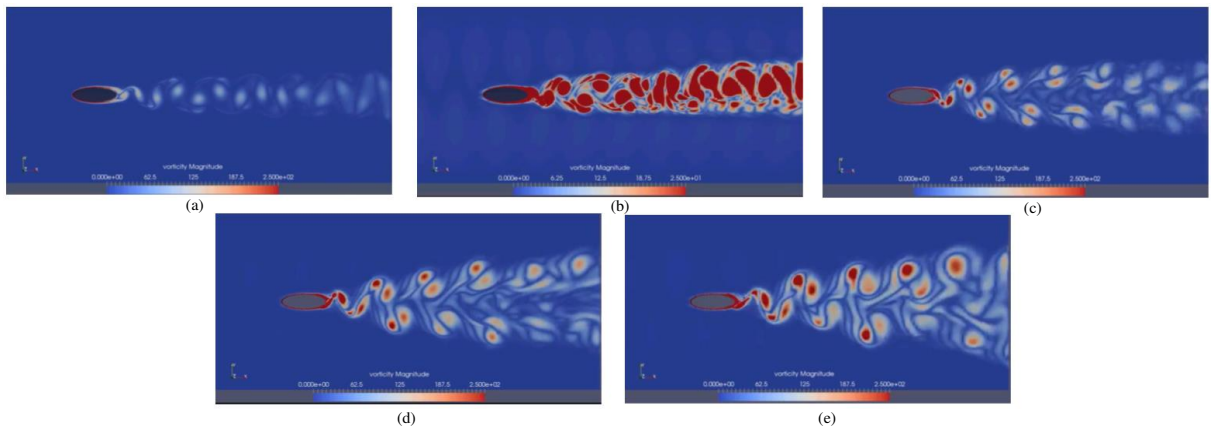

**Figure 5.** Vorticity modified design 2 at 1 m/s (a), 2 m/s (b), 3 m/s (c), 4 m/s (d), and 5 m/s (e) of wind speed

**Table 1.** Simulation results, oscillation frequency, lift coefficient, drag coefficient, and drift coefficient, on the bladeless wind turbine model

| V (m/s) | Frequency (Hz) | | | Lift Coefficient | | | Drag Coefficient | | | Drift Coefficient (CL/CD) | | |
|---|---|---|---|---|---|---|---|---|---|---|---|---|
| | ID | MD1 | MD2 | ID | MD1 | MD2 | ID | MD1 | MD2 | ID | MD1 | MD2 |
| 1 | 2,00 | 5,00 | 6,75 | 140,00 | 18,00 | 6,00 | 198,30 | 19,20 | 4,00 | 0,71 | 0,94 | 1,50 |
| 2 | 5,00 | 9,00 | 14,00 | 124,16 | 18,50 | 7,50 | 149,16 | 16,00 | 4,00 | 0,83 | 1,16 | 1,88 |
| 3 | 7,50 | 13,50 | 20,00 | 123,75 | 15,04 | 10,00 | 128,75 | 16,00 | 4,00 | 0,96 | 0,94 | 2,50 |
| 4 | 10,00 | 17,50 | 25,00 | 99,00 | 15,59 | 11,20 | 117,70 | 16,90 | 4,00 | 0,84 | 0,92 | 2,80 |
| 5 | 9,30 | 17,00 | 28,00 | 95,70 | 10,77 | 12,00 | 78,50 | 11,68 | 4,00 | 1,22 | 0,92 | 3,00 |

*ID: Initial design; MD1: Modified design 1; MD2: Modified design 2

Furthermore, the frequency data for each model is processed and produces a frequency graph for each model, which can be seen in Fig. 6.

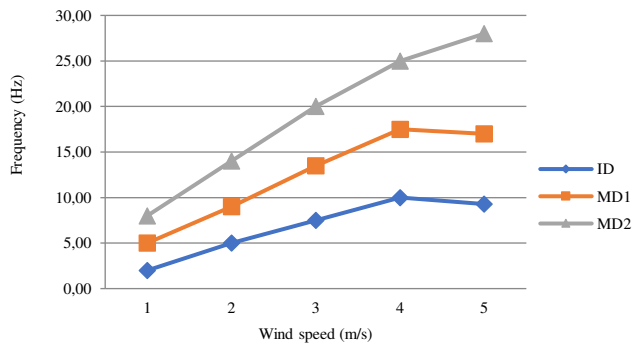

**Figure 6.** Vibration frequency of bladeless wind turbine mast to variation of mast model in each wind speed simulation

Based on Fig. 6, the MD2 model has a higher frequency value than the other 2 models, which is 8 Hz, followed by the MD1 model of 5 Hz and ID of 2 Hz at a wind speed of 1 m/s. In addition, the frequency value of each model at a wind speed of 5 m/s also shows that the MD2 model has the highest value of 28 Hz, followed by MD1 of 17 Hz and ID of 9.3 Hz. This shows that the MD2 model can oscillate well at wind speeds in the range of 1 m/s to 5 m/s compared to other models. In addition, a graph of the lift coefficient ratio (CL) with the drag coefficient (CD) is also obtained which is presented in Fig. 7.

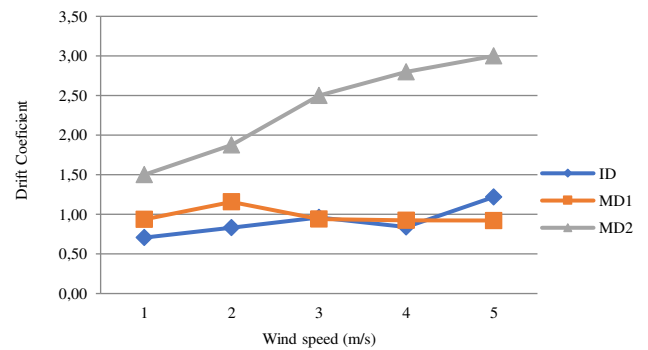

**Figure 7.** The value of the drift coefficient in each model at each wind speed

Based on Fig. 7, the ratio of the lift coefficient (CL) to the drag coefficient (CD) which produces the coefficient of oscillatory displacement force in the MD2 model has the largest CL/CD coefficient ratio compared to other models at each tested wind speed. So, it shows that the MD2 model has better sensitivity to wind motion than the ID and MD1 models at the same wind speed.

## 5. Conclusions

Based on the discussion, it can be concluded that: based on the stability of the fluid flow in the simulation results, the MD1 and MD2 models can already oscillate at a wind speed of 1 m/s, while ID begins to oscillate at a wind speed of 3 m/s; based on the air frequency on the vibration frequency graph, the MD2 model has the highest frequency value compared to the ID and MD1 models at each tested wind speed; and based on the ratio of lift coefficient (CL) to drag coefficient (CD) graph, the MD2 model has a large displacement coefficient or has better sensitivity to wind motion than other models.

MD2 in general has better results than MD1 and ID, and it can be said that MD2 can work more optimally in areas with low wind speeds, which is 1 to 5 m/s that suitable with an average wind speed in NTT, average wind speed 2.3 m/s. MD2 is a mechanical design model that is very suitable to be applied to NTT and the design can be expected to be one of the recommendations in making masts for BWT-based Wind Power Plants in Indonesia.

It is hoped that this research will be developed in the future using a 3-dimensional simulation method and with several other design models, so that it is expected to provide more valid findings regarding recommendations for mast mechanical design on BWT.

**Acknowledgements**

The authors would like to thank the Department of Nuclear Engineering and Physics Engineering UGM and all of those who contributed to acquiring research material.

**Data Availability**

The datasets generated during and/or analysed during the current study are available from the corresponding author on reasonable request.

**Declaration of Competing Interest**

The authors declare that they have no known competing financial interests or personal relationships that could have appeared to influence the work reported in this paper.